\documentclass[twocolumn,showpacs,preprintnumbers,amsmath,amssymb,showkeys]{revtex4}
\usepackage{graphicx}% Include figure files
\usepackage{dcolumn}% Align table columns on decimal point
\usepackage{bm}% bold math 

\begin{document}

\preprint{APS/123-QED}

\title{Spin-charge mixing effects on resonant tunneling in a polarized Luttinger Liquid}% Force line breaks with \\

\author{Kenji Kamide}
\email{kamide@kh.phys.waseda.ac.jp}
\author{Yuji Tsukada}
\author{Susumu Kurihara}
\affiliation{Department of physics, Waseda University, Okubo, Shinjuku, Tokyo 169-8555, Japan.}
\homepage{http://dyson.kh.phys.wasedas.ac.jp/}
\date{\today}% It is always \today, today,

\begin{abstract}
We investigate spin-charge mixing effect on resonant tunneling in spin-polarized Tomonaga-Luttinger liquid with double impurities. The mixing arises from Fermi velocity difference between two spin species due to Zeeman effect.
Zero bias conductance is calculated as a function of gate voltage $V_{\rm g}$, gate magnetic field $B_{\rm g}$, temperature and magnetic field applied to the system.
Mixing effect is  shown to cause rotation of the lattice pattern of the conductance peaks in $(V_{\rm g},B_{\rm g})$ plane, which can be observed in experiments.
At low temperatures, the contour shapes are classified into three types, reflecting the fact that effective barrier potential is renormalized towards ``perfect reflection'', ``perfect transmission'' and magnetic field induced ``spin-filtering'', respectively.
\end{abstract}

\maketitle

\section{\label{sec:level1}Introduction} 

Spin-charge separation and interaction dependent power laws of correlation functions have been known as non-Fermi-liquid behavior of Tomonaga-Luttinger (TL) liquid, which is expected to describe the low energy physics of one dimensional (1D) interacting electron systems \cite{gogolin}. 
In the past ten years, experimental evidences for the realization of TL states have been reported in many systems such as carbon nanotube devices \cite{nanotube transport1, nanotube transport2, postma}, quantum wires in semiconductor heterostructures~\cite{tarucha,yacoby1} and fractional quantum Hall systems~\cite{FQH1,FQH2}, as predicted by theories~\cite{egger&gogolin,wen}.
The power law temperature dependencies of the tunneling density of states~\cite{nanotube transport1,nanotube transport2,FQH1,FQH2}, and the bulk spectral density near the Fermi level~\cite{ishii} have been the key signatures of TL states in these experiments.
On the other hand, there are only a few experiments on the direct observation of spin-charge separation \cite{yacoby2,lee}.
 In order to obtain such low energy spectral profiles, we need to resolve local density of states at long distances from a scatterer or a boundary edge, by scanning tunneling microscope (STM)~\cite{lee}, which seems rather difficult to obtain high resolution data.
Instead, experiments on resonant tunneling \cite{postma,NTdot1,NTdot2} seems to be a more suitable procedure to detect the spin-charge separation, since they probes the energy level spacings in the quantum island. 

The resonant tunneling in TL liquid has been studied for more than a decade since Kane and Fisher's work \cite{kane-fisher,furusaki-nagaosa1,furusaki2}.
The charge transport is a main focus on those works.
Meanwhile, the spin transport in presence of external magnetic field, which is closely linked to detecting spin-charge separation, has not been studied enough.
There are some theoretical works concerned with this issue \cite{si1,egger2,balents1}.
They consider the quantum wire forms the quantum dot between ferromagnetic contacts.
There the spin and the charge transport can be controlled by the relative angle between the magnetization orientations of the ferromagnets \cite{egger2}.
In their results, the qualitative change due to the spin-charge separation can be seen.
However, the quantitative determination of the spin-charge separation in the excitation spectrum remains difficult. 

In this paper, we consider the resonant tunneling through double impurities in a spinful TL liquid under magnetic fields.
A schematic picture of the system is shown in Fig.\ref{fig:system}; a strong field $B$ is applied to the entire one-dimensional system.
 The charge and the spin in the region between double impurities are changed by a gate voltage $V_{\rm g}$ and a weak field $B_{\rm g}$, respectively.
Applying a strong magnetic field breaks the spin rotational symmetry and violate spin-charge separation in the spectral peaks \cite{si2}, due to the Fermi velocity differences between two spins.
As a result, the spin and the charge sector mix with each other~\cite{penc,kimura}.
It is shown that the spin-charge mixing effects can clearly be seen in  the resonant oscillation patterns of zero bias conductance in $(V_{\rm g}, B_{\rm g})$ plane.
The spin dependent scaling behavior of a single impurity potential, due to Zeeman effect, has been discussed by previous works~\cite{kimura,schmeltzer1,hikihara,lal}.
We also discuss a spin-charge mixing effects on the impurity scaling in the resonant tunneling.

 Zero bias conductance is calculated by standard bosonization technique~\cite{kane-fisher,furusaki-nagaosa1,furusaki2} as a function of a gate voltage, gate magnetic field, temperature $T$ and the strong magnetic field.
 The impurity potential $V(x)=V \left(\delta (x - d) + \delta(x+d)\right)$ is assumed to be either very weak ($V/v_{\rm F} \ll 1$) or very strong ($V/v_{\rm F} \gg 1$). 
%%%%%%%%%%%%%%%%%%%%%%%%%%%%%%%%%%
When $B \neq 0$, changing the gate voltage does affect the spin density together with the charge density due to the spin-charge mixing, leading to noticeable deformation in the lattice pattern of the conductance peaks~\cite{kamide}.
%%%%%%%%%%%%%%%%%%%%%%%%%%%%%%%%%%  
 Moreover contour shapes at low temperatures are divided into three types, depending on the bulk parameters such as $B$, interaction parameters $K_{\rho,s}$, where the impurity potentials are scaled towards ``perfect reflection", ``perfect transmission'' and magnetic field induced ``spin-filtering'', respectively.  
These three behaviors are explained by a renormalization group (RG) analysis of a single impurity potential in spin-charge mixed systems. 

\section{Tomonaga-Luttinger liquid under a magnetic field}
\begin{figure}[bp]
\includegraphics[width=78.1mm, height=28.8mm]{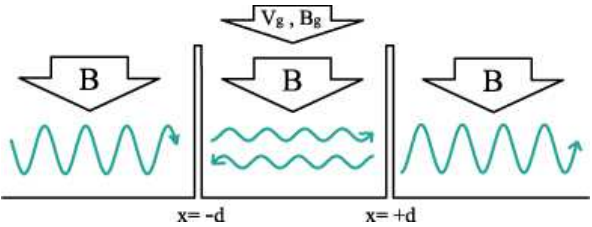}
\caption{\label{fig:epsart} Schematic figure of spin polarized Tomonaga-Luttinger liquid with two impurities under magnetic field. }
\label{fig:system}
\end{figure}

 We consider a system illustrated in Fig.\ref{fig:system}.
%%%%%%%%%%%%%%%%%%%%%%%%%%%
%%%%%%%%%%%%%%%%%%%%%%%%%%%
%%%%%%%%%%%%%%%%%%%%%%%%%%%%
%%%%%%%%%%%%%%%%%%%%%%%%
%%%%%%%%%%%%%%%%%%%%%%%%%%%
%%%%%%%%%%%%%%%%%%%%%%%%%%%% 
%%%%%%%%%%%%%%%%%%%%%%%%%%%%
%%%%%%%%%%%%%%%%%%%%%%%%
An infinite TL liquid, under a strong magnetic field $\overrightarrow{B}$ perpendicular to the wire, has two impurities (or barriers) of the strength $V$ at $x=\pm d$.
Hereafter we take the spin quantization axis in the direction of $\overrightarrow{B}$, and denote the strength of $\overrightarrow{B}$ by $B \equiv |\overrightarrow{B}|$.
$\hbar$ and $k_{\rm B}$ are set to unity in this paper.
Zeeman effect due to $B$ is incorporated into Hamiltonian as the difference in Fermi velocity between two spin species.
%%%%%%%%%%%%%%%%%%%%%%%%%%%%%%%%%%%%%%%%%%%%%%%%%%
For simplicity, we take into account the Zeeman effect only on spins with neglecting the orbital effects. This is allowed when the magnetic length $l_{\rm B}=\sqrt{1/eB}$ is longer than a width of the TL wire, as discussed in \cite{ajiki}.
%%%%%%%%%%%%%%%%%%%%%%%%%%%%%%%%%%%%%%%%%%%%%%%%%%
In the effective mass approximation $\varepsilon={k}^{2}/2m^{*}$, the velocity difference is given by, 
%%%%%%%%%%%%%%%%%%%%%%%%%%%%%%%%%%%%%%%%%%%%%%%%%%%%%%%%%%%%%%%%
\begin{eqnarray}
2\Delta 
%\equiv v_{\uparrow}-v_{\downarrow}
=v_{\rm F}(\sqrt{1+\frac{g \mu_{B} B}{2\varepsilon_{\rm F}}}-\sqrt{1-\frac{g \mu_{B} B}{2\varepsilon_{\rm F}}}).
\end{eqnarray}
%%%%%%%%%%%%%%%%%%%%%%%%%%%%%%%%%%%%%%%%%%%%%%%%%%%%%%%%%%%%%%%%
For not too strong fields, $\Delta$ is approximately linear in $B$.
For simplicity, we set $v_{\sigma}=v_{\rm F}+\sigma \Delta$, where the sign $\sigma=+(-)$ represents up (down) spin.
We shall call the region $-d<x<d$ an island. 
In the island, chemical potential of spin $\sigma $ electrons can be controlled by gate voltage $V_{{\text g}}$ and gate magnetic field $ B_{\rm g}$; $\delta \mu_{\sigma}=-e V_{{\text g}}+\sigma \frac{g \mu_{B}}{2} B_{\rm g}$.
We consider the situation where Zeeman energies due to strong magnetic field $B$ and due to gate magnetic field $B_{\rm g}$ are, respectively, on the order of Fermi energy and energy level spacing in the island; {\it i.e.}
$\varepsilon_{\rm F} \ge \frac{g \mu_{B} B}{2} \gg \frac{g \mu_{B} B_{\rm g} }{2} \sim  v_{\rm F} \frac{2 \pi}{2d} $.
Then the Fermi velocity change due to $B_{\rm g}$ can be neglected.

Hamiltonian of the system consists of four parts, 
\begin{eqnarray}
H \equiv H_{0}+H_{{\rm{int}}}+H_{{\rm{island}}}+H_{{\rm{b}}} \ , 
\end{eqnarray}
where $H_{0},H_{{\rm{int}}},H_{{\rm{island}}}$ and $H_{{\rm{b}}}$ are the Hamiltonian for free electrons, two-body interaction, electrons on the island and barrier potential, respectively. They are given in terms of fermion field operator,
\begin{subequations}
\label{eq:whole}
\begin{eqnarray}
&&H_{0}=\sum_{\sigma} \frac{v_{\sigma}}{i} \int dx  \left[ \psi^{\dagger}_{+,\sigma} \partial_{x} \psi_{+, \sigma}-\psi^{\dagger}_{-,\sigma} \partial_{x} \psi_{-, \sigma} \right] , \\
&&H_{\rm int}=\frac{1}{2} \sum_{\sigma,\sigma'} \int dxdy \ n_{\sigma}(x) {\mathcal U}_{\sigma,\sigma'}(x-y) n_{\sigma'}(y)   ,  \qquad \\
&&H_{{\rm island}}=  -eV_{\rm g}  \sum_{\sigma} \int_{-d}^{d} dx \ n_{\sigma}  \nonumber  \\
&& \qquad \quad +  \int_{-d}^{d} dx  \ \frac{1}{2} g \mu_{B} \overrightarrow{B_{\rm g}} \cdot ( \psi_{\alpha}^{\dagger} \vec{\sigma}_{\alpha,\beta}\psi_{\beta} )  ,   \label{h-island} \\
&&H_{{\rm b}}= V \sum_{\sigma} \ \left[ n_{\sigma}(-d) +  n_{\sigma}(d) \right] , \label{h-barriers}
\end{eqnarray} 
\end{subequations}
where $\psi_{\sigma} \equiv  \psi_{+,\sigma}e^{ik_{{\rm F},\sigma}x}+\psi_{-,\sigma} e^{-ik_{{\rm F},\sigma}x}$ and $n_{\sigma}=\psi^{\dagger}_{\sigma}\psi_{\sigma}$ are annihilation operator and density operator for electrons with spin $\sigma$. We neglect the charging energy of the island ($U$ in \cite{furusaki-nagaosa1}), which can be incorporated into $H_{\rm int}$.
$H_{0}+H_{\rm int}$ are written in terms of bosonic phase fields~\cite{gogolin},
%%%%%%%%%%%%%%%%%%%%%%%%%%%%%%%%%%%%%%%%%%%%%%
\begin{subequations}
\label{eq:whole}
\begin{eqnarray}
&&H_{0}+H_{\rm int}=\sum_{i=\rho, s} \frac{\pi v_{i}}{2} \int dx  
 \left[
 K_{i}^{-1} (\frac{1}{\pi}\partial_{x} \phi_{i})^2  + K_{i} \Pi_{i}^2
 \right] \nonumber
  \\
&& \qquad \qquad  +  \pi \Delta \int dx  \left[ (\frac{1}{\pi}\partial_{x} \phi_{\rho})(\frac{1}{\pi}\partial_{x} \phi_{s})+ \Pi_{\rho}\Pi_{s} \right],  \qquad
\label{hamiltonian before diagonalized}
\end{eqnarray}
\end{subequations}
%%%%%%%%%%%%%%%%%%%%%%%%%%%%%%%%%%%%%%%%%%%%%%
Here, $a$ is the inverse of the Fermi wave number:
$\phi_{i}$ and $\Pi_{i}$ are conjugate pairs of bosonic field with commutation relations $[\phi_{i}(x),\Pi_{j}(x')]=i \delta_{i,j} \delta(x-x')$, where 
$i=\rho$ $(s)$ represent for charge (spin) variables.
 
To diagonalize $H_{0}+H_{\rm int}$, we use the linear transformation~\cite{kimura} (see also Appendix B for the expression in ladder operators of TL bosons)
\begin{eqnarray}
%%%%%%%%%%%%%%%%%%%%%%%%%%%%%%%%%%%%%%%%%
\left(
\begin{array}{c}
\phi_{\rho}  \\
\phi_{s}
\end{array}
\right)
=
\left(
\begin{array}{cc}
\cos \alpha & - \frac{1}{y}\sin \alpha \\
y \sin \alpha & \cos \alpha
\end{array}
\right)
\left(
\begin{array}{c}
\tilde{\phi}_{\rho}  \\
\tilde{\phi}_{s}
\end{array}
\right), \nonumber \\
%%%%%%%%%%%%%%%%%%%%%%%%%%%%%%%%%%%%%%%%%%%
\left(
\begin{array}{c}
\Pi_{\rho}  \\
\Pi_{s}
\end{array}
\right)
=
\left(
\begin{array}{cc}
\cos \alpha & -y \sin \alpha \\
\frac{1}{y}\sin \alpha & \cos \alpha
\end{array}
\right)
\left(
\begin{array}{c}
\tilde{\Pi}_{\rho}  \\
\tilde{\Pi}_{s}
\end{array}
\right).
\label{linear transform}
\end{eqnarray}
Commutation relations are preserved under this transformation $[\tilde{\phi}_{i}(x),\tilde{\Pi}_{j}(x')]=i \delta_{i,j} \delta(x-x')$. Parameters $\alpha$ and $y$ are, respectively, a rotation angle and a scale factor in ``spin-charge space'' given by,
\begin{eqnarray} 
y=\sqrt{
\frac{v_{\rho}K_{\rho}^{-1}+v_{s}K_{s}}{v_{\rho}K_{\rho}+v_{s}K_{s}^{-1}}
}, \ 
\tan 2\alpha =\frac{2\Delta}{v_{\rho}K_{\rho}y-v_{s}K_{s}y^{-1}}. \label{rotation angle} 
\end{eqnarray} 
The rotation angle $\alpha$ is proportional to $\Delta$, and thus to $B$, for small $\Delta$.
After the transformation, $H_{\rm TL}\equiv H_{0}+H_{\rm int}$ becomes,
%%%%%%%%%%%%%%%%%%%%%%%%%%%%%%%%%%%%%%%%%%%%%%
\begin{eqnarray}
H_{\rm TL}=\sum_{i=\rho, s} \frac{\pi \tilde{v}_{i}}{2} \int dx \left[ 
 \tilde{K}_{i}^{-1} (\frac{1}{\pi}\partial_{x} \tilde{\phi}_{i})^2  + \tilde{K}_{i} \tilde{\Pi}_{i}^2 \right] .
 \label{TL hamiltonian}
\end{eqnarray}
%%%%%%%%%%%%%%%%%%%%%%%%%%%%%%%%%%%%%%%%%%%%%%
The expression of $\tilde{v}_{i}$ and $\tilde{K}_{i}$ are given in Appendix A.

As discussed in \cite{kimura,schmeltzer1,hikihara,lal}, in a polarized TL liquid, the scaling dimensions of a single impurity potential split between two spins.
Renormalization group equations for small backscattering amplitude $V_{\sigma}$ and small tunneling amplitude $t_{\sigma}$ are given
%%%%%%%%%%%%%%%%%%%%%%%%%%%%%%%%%%%%%%%%%%%%%%
\begin{eqnarray}
\frac{d V_{\sigma}}{dl}&=&(1-\frac{\eta_{\sigma}}{2}) V_{\sigma} \quad {\rm for} \quad V_{\sigma} \ll v_{\rm F},  \label{eq: single impurity scaling (weak)} \\
\frac{d t_{\sigma}}{dl}&=&(1-\frac{\lambda_{\sigma}}{2}) t_{\sigma} \quad {\rm for} \quad V_{\sigma} \gg v_{\rm F}, \label{eq: single impurity scaling (strong)}
\end{eqnarray}
%%%%%%%%%%%%%%%%%%%%%%%%%%%%%%%%%%%%%%%%%%%%%%
with $l=\ln{\Lambda/\Lambda'}$ and an initial (running) energy cutoff $\Lambda$ ($\Lambda'$).
The scaling dimensions $\eta_{\sigma}$ and $\lambda_{\sigma}$ are given
%%%%%%%%%%%%%%%%%%%%%%%%%%%%%%%%%%%%%%%%%%%%%%
\begin{eqnarray}
\eta_{\sigma}&=&\tilde{K}_{\rho} B_{\rho,\sigma}^2 + \tilde{K}_{s} B_{s,\sigma}^2 ,
\label{scaling dimension of impurity backscattering}  \\ 
\lambda_{\sigma}&=&\tilde{K}_{\rho}^{-1}D_{\rho,\sigma}^{2}+\tilde{K}_{s}^{-1}D_{s,\sigma}^{2}. \label{eq:boundary exponent}
\end{eqnarray}
%%%%%%%%%%%%%%%%%%%%%%%%%%%%%%%%%%%%%%%%%%%%%%
where $B_{\rho,\sigma}= \cos{\alpha} +\sigma y \sin{\alpha}$, $B_{s,\sigma}=- y^{-1} \sin{\alpha}+\sigma \cos{\alpha}$, $D_{\rho,\sigma}=\cos{\alpha}+\sigma y^{-1}\sin{\alpha}$, and $D_{s,\sigma}=\sigma \cos{\alpha}- y\sin{\alpha}$.
Thus the ratio of reflection amplitudes and tunneling amplitudes scale like, respectively, $V_{\uparrow}/V_{\downarrow} \propto (T/\Lambda)^{\frac{\eta_{\uparrow}-\eta_{\downarrow}}{2}}$ and $t_{\uparrow}/t_{\downarrow} \propto (T/\Lambda)^{\frac{\lambda_{\uparrow}-\lambda_{\downarrow}}{2}}$.
The difference of the exponents $\delta \eta= (\eta_{\uparrow}-\eta_{\downarrow})/2$ and $\delta \lambda= (\lambda_{\uparrow}-\lambda_{\downarrow})/2$ are given
%%%%%%%%%%%%%%%%%%%%%%%%%%%%%%%%%%%%%%%%%%%%%%%%%%%%%%%%%%%%%%%%%%%%%%%
\begin{eqnarray}
\delta \eta  &=&(\tilde{K}_{\rho}y-\tilde{K}_{s}y^{-1})\sin{2\alpha} , \\
\delta \lambda &=& \frac{\tilde{K}_{s}y^{-1}-\tilde{K}_{\rho}y}{\tilde{K}_{\rho} \tilde{K}_{s}}\sin{2\alpha}. 
\end{eqnarray}
%%%%%%%%%%%%%%%%%%%%%%%%%%%%%%%%%%%%%%%%%%%%%%%%%%%%%%%%%%%%%%%%%%%%%%%%%
This expression tells us that the split of the scaling dimensions between two spins become large when the spin-charge mixing angle is large.
Due to the split of exponents, we expect the spin current with large polarization $P=\frac{t_{\uparrow}-t_{\downarrow}}{t_{\uparrow}+t_{\downarrow}}$ at low temperature in the strong barrier limit.
This scaling effect with the exponents $\eta_{\sigma}$ and $\lambda_{\sigma}$ also appears in the conductance for double barrier structure.
%%%%%%%%
\section{Weak barrier limit ($V \to 0$)}
%%%%%%%%
We consider first the weak barrier limit $V \to 0$, where the electron transfer is due to the coherent tunneling of the spin-charge mixed density wave.
The zero bias conductance is calculated perturbatively with respect to the small barrier potential $\frac{V}{\pi a}$.
The Hamiltonian for the island and for the barriers in Eq.(\ref{h-island}) and Eq.(\ref{h-barriers}) are written in terms of the bosonic phase at $x=\pm d$,
\begin{eqnarray}
&& H_{\rm{island}}= \frac{-2eV_{g}}{\pi}\theta_{\rho}^{-}
+\frac{g \mu_{B} B_{g}}{\pi}  \theta_{s}^{-}  , \\
&&H_{\rm{b}}=\frac{V}{\pi a} \sum_{\sigma=\pm} \sin(\theta_{\rho}^{+}+\sigma \theta_{s}^{+}) \cos(2k_{{\rm F},\sigma}d +\theta_{\rho}^{-}+\sigma \theta_{s}^{-}). \nonumber 
\\
&& 
\end{eqnarray}
 $\theta_{i}^{\pm}=(\phi_{i}(d) \pm \phi_{i}(-d))/\sqrt{2}$ is the linear combination of phases at $x=\pm d$.
 According to our assumption that the magnetic field $B$ is strong enough that $(k_{\rm F,\uparrow}-k_{\rm F,\downarrow})d \gg \pi$, only $z$-component of $\overrightarrow{B_{\rm g}}$ remains after the integration of (\ref{h-island}) with neglecting the fast oscillating terms. Hereafter, we denote $( \overrightarrow{B_{\rm g}} )_{\rm z}$ by $B_{\rm g}$. 
%%%%%%%%

To make calculations easier, it is helpful to construct an effective action \cite{kane-fisher} obtained by integrating out the TL field except for the positions of the barriers $x=\pm d$, since charge (spin) current through barriers depends only on the local variables, $j_{\rho (s)}=(2/\pi)\partial_{t} \theta_{\rho(s)}^{+}$. 
The integrated effective action is calculated as,
%%%%%%%%%%%%%%%%%%%%%%%%%%%%%%%%%%%%%%%%%%%%%%%
\begin{eqnarray}
&&S_{\rm{eff}}=\sum_{
\omega_{n},j=\pm,i=\rho,s}
\tilde{\epsilon}_{i}^{j}(\omega_{n})\tilde{\theta}_{i}^{j}(\omega_{n})\tilde{\theta}_{i}^{j}(-\omega_{n}) 
\nonumber \\
&& \qquad \qquad
 +\int_{0}^{\beta}d\tau \left[ H_{{\rm island}}+ H_{{\rm b}}\right], \\
&& \tilde{\epsilon}_{i}^{\pm}(\omega_{n}) = \frac{1}{\pi \tilde{K}_{i}}\frac{| \omega_{n}|}{1 \pm \exp \{-(2d/\tilde{v}_{i})|\omega_{n}| \}}. 
\end{eqnarray}
%%%%%%%%%%%%%%%%%%%%%%%%%%%%%%%%%%%%%%%%%%%%%%%
Here we have defined new variables $\tilde{\theta}_{i}^{\pm} \equiv (\tilde{\phi}_{i}(d) \pm \tilde{\phi}_{i}(-d))/\sqrt{2}$. 
$\tilde{\theta}$ and $\theta$ relates each other by the same linear transformation between $\tilde{\phi}$ and $\phi$.
We will calculate zero bias conductance for charge (spin) current following Kubo formula,
\begin{equation}
G_{\rho (s)}=\lim_{\omega_{n} \to 0} \frac{2e^2}{\omega_{n}} \left< j_{\rho (s)}(\omega_{n}) j_{\rho}(-\omega_{n}) \right>, \nonumber 
\end{equation}
where analytic continuation is to be taken before taking the limit.
The effective action $S_{{\rm eff}}$ is used for taking thermal average such that
\begin{eqnarray}
\left< A \right>=\frac{\int {\prod_{i,j} D\tilde{\theta}_{i}^{j}}  A \exp\{-S_{{\rm eff}}\}}
{\int \prod_{i,j} D\tilde{\theta}_{i}^{j}  \exp \{-S_{{\rm eff}}\} } \ .
\end{eqnarray}
Hamiltonian for the island $H_{\rm island}$ is given after canonical transformation (\ref{linear transform}) as
\begin{eqnarray}
 H_{{\rm island}}&=&
 A_{1}\tilde{\theta}_{\rho}^{-} +A_{2}\tilde{\theta}_{s}^{-}, 
\end{eqnarray}
with the coefficients,
\begin{eqnarray} 
&& \ 
A_{1}=-\frac{2eV_{g}}{\pi} \cos{\alpha} + \frac{g\mu_{B}B_{\rm g}}{\pi}y \sin{\alpha}, \nonumber \\ 
&& \ A_{2}=\frac{2eV_{g}}{\pi} \frac{1}{y}\sin{\alpha} + \frac{g\mu_{B}B_{\rm g}}{\pi} \cos{\alpha}. \nonumber 
\end{eqnarray}
Thus, the terms other than barrier potentials, quadratic in $\tilde{\theta}_{i}^{\pm}$, can be treated exactly in $S_{{\rm eff}}$.

 After straightforward calculation, we obtain an explicit form of the conductance to the second order in $\frac{V}{\pi a}$; $G_{i} = G_{i}^{(0)}+G_{i}^{(2)}$ for $i=\rho , s$.
The unperturbed conductance $G_{\rho}^{(0)}$ and $G_{s}^{(0)}$ are \cite{kimura}, 
\begin{eqnarray}
G_{\rho}^{(0)}&=&\frac{e^2}{\pi}(\tilde{K}_{\rho}{\cos^{2} {\alpha}} +\tilde{K}_{s} y^{-2}\sin^{2}\alpha  ), \label{eq: conductance in clean limit 1} \\
G_{s}^{(0)}&=&\frac{e^2}{2\pi}(\tilde{K}_{\rho}y-\tilde{K}_{s} y^{-1}) \sin{2\alpha} \ . \label{eq: conductance in clean limit 2}
\end{eqnarray}
%%%%%%%%%%%%%%%%%%%%%%%%%%%%%%%%%%%%%%%%%%%%%%%%%%%%%%%%%%%%%%%%%%%%%%
This expression implies that spin polarized current can flow in a clean, infinite TL liquid due to spin-charge mixing effect, since  $G_{s}^{(0)}=0$ when $\Delta =0$ for all interaction parameters. 
However, such a violation of the conductance quantization is shown to be an artifact for infinite system as previous works pointed out for unpolarized system~\cite{ref12,ref13,ref14}.
We show in Appendix C that the prefactors of conductance should vanish $i.e.$ $G_{\uparrow}^{(0)}=G_{\downarrow}^{(0)}=\frac{e^2}{2 \pi}$ also for spin polarized system, by taking into account the effects of Fermi liquid reservoirs.
Thus the spin polarized current cannot be generated in a clean system, even when a magnetic field are applied.

The second order correction $G_{i}^{(2)}$ is
%%%%%%%%%%%%%%%%%%%%%%%%%%%%%%%%%%%%%%%%%%%%%%%%%%%%%%%%%%%%%%%%%%%%%%%%
\begin{widetext}
\begin{subequations}
\label{eq:whole}
\begin{eqnarray}
&&G_{i}^{(2)} = -\frac{e^2}{4}
\left( \frac{V}{\pi a} \right)^2 \sum_{\sigma}
%%%%%%%%%%%%%%%%%%%%%%%%%%%%%%
C_{\sigma}^{i} \lim_{\omega_{n} \to 0}
\int_{0}^{\beta}d\tau \ \frac{1-\cos{\omega_{n}\tau}}{\omega_{n}}R_{\sigma}
\exp
\bigl[
-\frac{1}{2\beta}
\sum_{\omega'_{n}} 
\bigl( 
\frac{B_{\rho,\sigma}^2}{\tilde{\epsilon}_{\rho}^{+}(\omega'_{n})}
+
\frac{B_{s,\sigma}^2}{\tilde{\epsilon}_{s}^{+}(\omega'_{n})}
\bigr)
(1-\cos{\omega'_{n}\tau})
\bigr] , \\ 
&& \qquad R_{\sigma}
= \cos{\Omega_{\sigma}} 
\exp\left[-\sum_{\omega'_{n}}f_{\sigma}(\omega_{n}' ) (1+\cos{\omega'_{n}\tau})
\right]
+\exp\left[-\sum_{\omega'_{n}}f_{\sigma}(\omega_{n}' ) (1-\cos{\omega'_{n}\tau})
\right] \\ 
&& \qquad \Omega_{\sigma}=-4k_{{\rm F},\sigma}d+
\bigl[ \frac{-\cos{\alpha}B_{\rho,\sigma}}{\pi \tilde{\epsilon}_{\rho}^{-}(0)}
+\frac{y^{-1}\sin{\alpha} B_{s,\sigma}}{\pi \tilde{\epsilon}_{s}^{-}(0)} 
\bigr]
2eV_{g}
+ \bigl[ \frac{y \sin{\alpha}B_{\rho,\sigma}}{\pi \tilde{\epsilon}_{\rho}^{-}(0)}
+\frac{\cos{\alpha}B_{s,\sigma}}{\pi \tilde{\epsilon}_{s}^{-}(0)}
\bigr]
g \mu_{B}B_{g} , \\
&& \qquad
f_{\sigma}(\omega_{n}' )=
\frac{1}{2\beta} 
\left(
\frac{B_{\rho,\sigma}^2 }{ \tilde{\epsilon}_{\rho}^{-}(\omega'_{n})}
+
\frac{B_{s,\sigma}^2 }{\tilde{\epsilon}_{s}^{-}(\omega'_{n})}
\right) , 
\end{eqnarray}
\end{subequations}
\end{widetext}
%%%%%%%%%%%%%%%%%%%%%%%%%%%%%%%%%%%%%%%%%%%%%%%%%%%%%%%%%%%%%%%%%%%
with $\beta=1/T$ and the constants given by,
%%%%%%%%%%%%%%%%%%%%%%%%%%%%%%%%%%%%%%%%%%%%%%%%%%%%%%%%%%%%%%%%%%%%
\begin{eqnarray}
&& C_{\sigma}^{\rho}=\left( \cos{\alpha} \ B_{\rho,\sigma} \tilde{K}_{\rho}
-y^{-1}\sin{\alpha} \  B_{s,\sigma} \tilde{K}_{s} \right)^2  , \nonumber \\
&& C_{\sigma}^{s}=\left( \cos{\alpha} \ B_{\rho,\sigma} \tilde{K}_{\rho}
-y^{-1}\sin{\alpha} \  B_{s,\sigma} \tilde{K}_{s} \right) \nonumber \\
&& 
\times \left( y \sin{\alpha} \ B_{\rho,\sigma} \tilde{K}_{\rho}
+ \cos{\alpha} \  B_{s,\sigma} \tilde{K}_{s} \right). \nonumber
\end{eqnarray}
The second order correction $G_{i}^{(2)}$ gives us some information about the resonant oscillation in $(V_{g},B_{g})$ plane;
the exponential terms determine the temperature dependent amplitude, and $\Omega_{\sigma}$ determines the period.
As seen in the above expression, the conductance oscillation comes from a sum of two independent contributions;
 resonant tunneling of spin up electrons and spin down electrons.
This is because the average of a product of cosine terms for up spin and down spin always vanishes; 
$\left< 
\sin \bigl(\theta_{\rho}^{+}(\tau)+\theta_{s}^{+}(\tau) \bigr)
\sin \bigl(\theta_{\rho}^{+}(0)-\theta_{s}^{+}(0) \bigr)
 \right>=0$.

At low temperature $T \ll \tilde{\epsilon}_{\rho}^{-}(0),\tilde{\epsilon}_{s}^{-}(0)$, it is given
\begin{eqnarray}
G_{i}^{(2)}=-\frac{e^2}{2\pi}
\bigl( \frac{V}{a \Lambda} \bigr)^{2}
\sum_{\sigma} C_{\sigma}^{i}
\frac{\Gamma(\frac{\eta_{\sigma}}{2})}{\Gamma(\frac{\eta_{\sigma}+1}{2})/\Gamma(\frac{3}{2})} \nonumber \\
\times (1+\cos{\Omega_{\sigma}})
\bigl(\frac{\pi T}{\Lambda}
\bigr)^{\eta_{\sigma}-2}, \label{eq: weak barrier conductance}
\end{eqnarray}
where $\Lambda$ is the high energy cut-off.
At high temperature $T \gg \tilde{\epsilon}_{\rho}^{-}(0),\tilde{\epsilon}_{s}^{-}(0)$, the term proportional to $\cos{\Omega_{\sigma}}$ is exponentially suppressed as increasing temperature, while the other terms remain unchanged in (\ref{eq: weak barrier conductance}).
Thus, the conductance oscillation dessappers and the double impurity structure can be viewed as a single impurity whose scaling law is given in Eq.(\ref{eq: single impurity scaling (weak)}) for $\frac{\pi v_{\rm F}}{2 d} \ll T \ll \Lambda$. 
%%%%%%%%%%%%%%%%%%%%%%%%%%%%%%%%%%%%%%%%%%%%%%%%%%%%%%%%%%%%%%%%%%%%%%%%%%%%%%%
\begin{figure}[hbp]
\includegraphics[width=78.48mm, height=81.84mm]{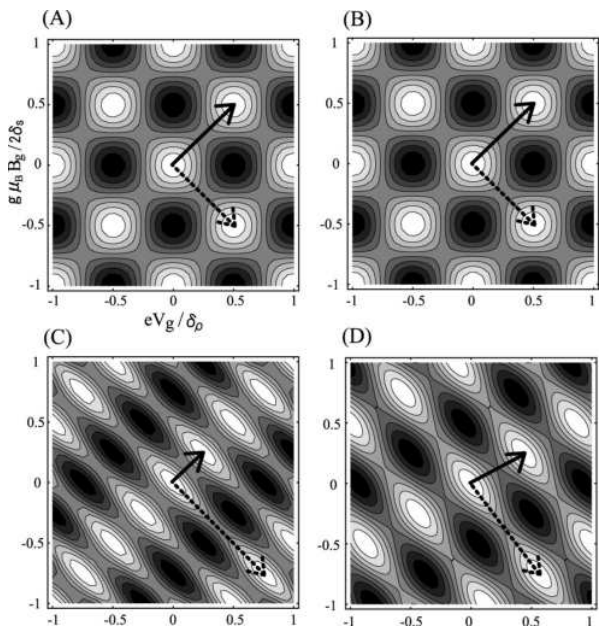}
\caption{\label{fig:epsart} Second order conductance correction $G_{\rho}^{(2)}$ are plotted as a function of gate voltage and magnetic field normalized by $\delta_{i}=\frac{v_{i}}{K_{i}} \frac{\pi }{2d}$ for Hubbard model.
$K_{\rho}=v_{\rm F}/v_{\rho}$, $K_{s}=v_{\rm F}/v_{s}=1$; 
(A) $K_{\rho}=1, \ \Delta/v_{\rm F}=0$, (B) $K_{\rho}=0.5, \ \Delta/v_{\rm F}=0$, 
(C) $K_{\rho}=1, \ \Delta/v_{\rm F}=0.5$, (D) $K_{\rho}=0.5, \ \Delta/v_{\rm F}=0.5$. 
The temperature is fixed at $T/\Lambda=0.1$ in all figures.
Solid (broken) arrow indicates the direction of peak line of the conductance deviation due to up (down) spin resonant tunneling. }
\label{fig:conductance}
\end{figure}
%%%%%%%%%%%%%%%%%%%%%%%%%%%%%%%%%%%%%%%%%%%%%
%%%%%%%%%%%%%%%%%%%%%%%%%%%%%%%%%%%%%%%%%%%%%%%%%%%%%%%%%%%%%%%%%%%%%%%%%%%

Figure \ref{fig:conductance} shows the contour plot of the conductance correction as a function of gate voltage and gate magnetic field.
Conductance peaks form lattice in ($V_{\text g} \ ,B_{\text g}$) plane and spin-charge mixing effect causes the deformation of the lattice.
%%%%%%%%%%%%%%%%%%%%%%%%%%%%%%%%%%%%%%%%%%%%%%%%%%%%%%%%%%%%%%%%%%%%%%
For unpolarized Fermi liquid $\Delta =0$ and $K_{\rho}=K_{s}=1$ in Fig.\ref{fig:conductance}(A), conductance peaks form rectangular lattice with periods $( eV_{\rm g})_{0}=(\frac{1}{2}\mu_{\rm B} B_{\rm g})_{0}=\frac{\pi v_{\rm F}}{2d}$.
Changing the interaction parameter from $K_{\rho}=1$ to $K_{\rho}=0.5$ (from Fig.\ref{fig:conductance}(A) to Fig.\ref{fig:conductance}(B)), the lattice shape does not change but with a shift in the period $( eV_{\rm g})_{0}=\frac{\pi v_{\rho}}{2dK_{\rho}}$ due to the change in charge- and spin- susceptibility of the island $\delta_{\rho}=\frac{\pi v_{\rho}}{2dK_{\rho}}$ and $\delta_{s}=\frac{\pi v_{s}}{2dK_{s}}$.
As increasing $\Delta$ from $\Delta=0$ to $\Delta=0.5 v_{\rm}$ in Fermi liquid case (from Fi.\ref{fig:conductance}(A) to Fig.\ref{fig:conductance}(C)), stretches in primary unit vectors of the lattice (arrows in figures), by a factor $(\sim 1+\sigma \frac{\Delta}{v_{\rm F}})$ for spin $\sigma$ electrons, are seen.
It reflects the change in the level spacing due to Zeeman effect given by $\sigma \frac{2\pi \Delta}{2d}$.
Then, adding the interaction from $K_{\rho}=1$ to $K_{\rho}=0.5$ (from Fig.\ref{fig:conductance}(C) to Fig.\ref{fig:conductance}(D)), rotations of the primary unit vectors rotates reflecting rotations of spin-charge space.
This is exactly due to spin-charge mixing effect.
For $K_{\rho}=K_{s}$, the rotation cannot be seen in Fig.\ref{fig:conductance}(C) since the principal axis in spin-charge space stay in the direction $\pm \frac{\pi}{4}$ independent of $\Delta$.
%%%%%%%%%%%%%%%%%%%%%%%%%%%%%%%%%%%%%%%%%%%%%%%%%%%%%%%%%%%%%%%%%%%%%%%
The rotation angle $\delta \vartheta$ is proportional to ``spin-charge mixing angle $\alpha$'' for small $\Delta$,
\begin{eqnarray}
\delta \vartheta =\frac{1}{2} \bigl(
y+y^{-1}-y\frac{v_{s}/K_{s}}{v_{\rho}/K_{\rho}}- {y^{-1}} \frac{v_{\rho}/K_{\rho}}{v_{s}/K_{s}} \bigr) \alpha +{\mathcal O}( \alpha^2 ), \nonumber \\
\end{eqnarray}
%%%%%%%%%%%%%%%%%%%%%%%%%%%%%%%%%%%%%%%%%%%%%%%%%%%%%%%%%%%%%%%%%%%%%%
when there are only density-density interactions between electrons $H_{\rm int}\propto g_{\rho}(\frac{1}{\pi}\partial_{x} \phi_{\rho})^2 +g_{s}(\frac{1}{\pi}\partial_{x} \phi_{s})^2$ {\it i.e.} TL parameters satisfy $K_{\rho}v_{\rho}=K_{s}v_{s}=v_{\rm F}$,
%%%%%%%%%%%%%%%%%%%%%%%%%%%%%%%%%%%%%%%%%%%%%%%%%%%%%%%%%%%%%%%%%%%%%%
\begin{eqnarray}
\delta \vartheta =\frac{K_{\rho}^{2}-K_{s}^{2}}{2} \frac{\Delta}{v_{\rm F}} +{\mathcal O} \left( (\Delta / v_{\rm F})^2 \right). \label{eq: rotation angle}
\end{eqnarray}
%%%%%%%%%%%%%%%%%%%%%%%%%%%%%%%%%%%%%%%%%%%%%%%%%%%%%%%%%%%%%%%%%%%%%%

\begin{figure}[bp]
\includegraphics[width=76.48mm, height=39.68mm]{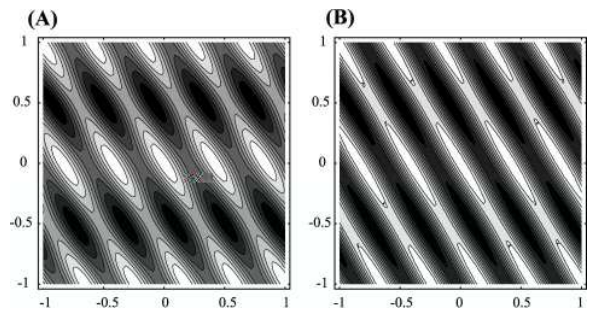}
\caption{\label{fig:epsart} $G_{\rho}^{(2)}$ are plotted as a function of gate voltage and magnetic field normalized by $\delta_{i}=\frac{v_{i}}{K_{i}} \frac{\pi }{2d}$ for spin anisotropic interaction parameters.
$K_{\rho}=v_{\rm F}/v_{\rho}=0.6$, $K_{s}=v_{\rm F}/v_{s}=1.4$, $\Delta/v_{\rm F}=0.5$; 
(A) $T/\Lambda=10^{-2}$, 
(B) $T/\Lambda=10^{-6}$.}
\label{fig:conductance-low temperature}
\end{figure}

When the temperature decreases, the amplitude of oscillation for two spins follow power laws with different exponents.
In case of spin isotropic interaction ($K_{s}=1$ $i.e.$ $g_{s}=0$), depending on the interaction $g_{\rho}=\frac{1}{2}(K_{\rho}^{-2}-1)$ is repulsive or attractive, impurity potentials are scaled toward perfect reflection or perfect transmission.
Extending the parameter space to spin anisotropic interaction ($K_{s} \neq 1$ $i.e.$ $g_{s} \neq 0$),  ``spin filter phase'' where $\eta_{\uparrow}>2, \eta_{\downarrow}<2$, emerges between the two phases ``perfect transmission'' and ``perfect reflection''. 
In this phase, the impurity potential scales toward perfect transmission for up spin (majority spin), and perfect reflection for down (minority) spin.
The conductance correction at different temperatures are plotted with TL parameters in spin filter phase in Fig.\ref{fig:conductance-low temperature}.
The peaks of the conductance correction changes from lattice structure to one plane wave, as lowering temperature from Fig.\ref{fig:conductance-low temperature}(A) to Fig.\ref{fig:conductance-low temperature}(B).
This is due to the strong suppression of backscattering current of the majority spins, and the enhansment for the minority spin at low temperature.
These results suggest the possibility of generating and modulating spin current with large polarization by gate voltage and gate magnetic field.
%%%%%%%%%%%%%%%%%%%%%%%%%%%%%%%%%%%%%%%%%%%%%%%%%%%%%%%%%%%%%%%%%%%%%%%%%%%%%%
%%%%%%%%%%%%%%%%%%%%%%%%%%%%%%%%%%%%%%%%%%%%%%%%%%%%%%%%%%%%%%%%%%%%%%%%%%%%%%

\section{Weak link limit ($V\to \infty$)}
In this section, we consider the weak link limit ($V \to \infty$), where the tunneling amplitudes $t_{\sigma}$ satisfies $t_{\sigma}/v_{\rm F} \ll 1$ and the electron transport is described by a sequential tunneling picture~\cite{furusaki-nagaosa1,furusaki2}.
We will calculate zero-bias conductance perturbatively to the lowest order within a master equation framework.
The higher order contributions, such as cotunneling (COT)~\cite{furusaki-nagaosa1,furusaki2} and correlated sequential tunneling (CST)~\cite{thorwart}, are neglected.
Contributions from COT and CST become important, respectively, away from the resonance peaks and for rather transparent barriers~\cite{thorwart}.
Thus, they do not seem to make major changes in the discussions here focusing on the peak positions and the height in the weak link limit.

We consider a quantum island with length $2d$ weakly linked to the semiinfinite TL wires at the both ends. The Hamiltonian for the semiinfinite wires $H_{\rm wires}$ are the same as Eq.(\ref{TL hamiltonian}). We represent electron field operators of the island, the right (R) and the left (L) wire by $\psi_{\sigma}=\sum_{\tau} \psi_{\tau,\sigma}$, $\psi_{\sigma}^{\rm R}=\sum_{\tau} \psi_{\tau,\sigma}^{\rm R}$ and $\psi_{\sigma}^{\rm L}=\sum_{\tau} \psi_{\tau,\sigma}^{\rm L}$, respectively.
They obey the open boundary condition $\psi_{\sigma}(\pm d)=\psi_{\sigma}^{\rm R}(d)=\psi_{\sigma}^{\rm L}(-d)=0$. The chemical potentials of the right and the left wire are set $\mu_{\rm R}=-eV_{\rm bias}$ and $\mu_{\rm L}=0$, and we take the limit $ eV_{\rm bias} \to 0$ at the last stage of the calculations for the zero bias conductance.
In the sequential tunneling regime, the number of excess electrons on the island $N_{\sigma}$ becomes a good quantum number, which enters in the zero mode Hamiltonian $H_{\rm zero}=H_{\rm d}+H_{\rm island}$,
%%%%%%%%%%%%%%%%%%%%%%%%%%%%%%%%%%%%%%%%%%%%%%%%%%%%%%%%%%%%%%%%%%%%%%%%%%%%%%
\begin{eqnarray}
&&H_{\rm d}=\frac{\pi v_{\rm F}}{4d}\sum_{\sigma}
\left(
\frac{K_{\rho}^{-2}+K_{s}^{-2}}{2}+\sigma \frac{\Delta}{v_{\rm F}}
\right) N_{\sigma}^{2} \nonumber  \\
&&\qquad \quad +\frac{\pi v_{\rm F}}{2d} 
\left(
\frac{K_{\rho}^{-2}-K_{s}^{-2}}{2}
 \right)
 N_{\uparrow}N_{\downarrow}, \label{eq:zero mode 1} \\
&& H_{\rm island}= \sum_{\sigma} 
 \left( -eV_{g}
 +\sigma \frac{g\mu_{B}B_{g}}{2}
 \right)
 N_{\sigma}. \label{eq:zero mode 2}
\end{eqnarray}
%%%%%%%%%%%%%%%%%%%%%%%%%%%%%%%%%%%%%%%%%%%%%%%%%%%%%%%%%%%%%%%%%%%%%%%%%%%%%%
The Hamiltonian for non-zero mode $H_{\rm fluc}$ is the same as in Eq.(\ref{TL hamiltonian}); $H_{\rm fluc}=\sum_{i=\rho, s ,n>0} \tilde{v}_{i}k_{n}(\tilde{\alpha}_{i,n}^{ \dagger}\tilde{\alpha}_{i,n}+1/2)$, where $k_{n}=\pi n /2d$. 
Under the boundary condition, the mode expansion of the electron field operator $\psi_{\sigma}$ becomes~\cite{fabrizio1,mattsson,a-fabrizio,schmeltzer1},
\begin{eqnarray}
&& \psi_{\tau,\sigma}=\sqrt{\frac{k_{{\rm F},\sigma}}{\pi}}e^{i\tau k_{{\rm F},\sigma}(x+d) +i\tau(\chi_{\tau,\sigma}^{0} + \chi_{\tau,\sigma})}, \\
&& \chi_{\tau,\sigma}^{0}=\frac{\pi}{2}+\frac{\pi N_{\sigma}(x+d)}{2d}+\tau \theta_{\sigma}, \qquad \\
&&  \chi_{\tau,\sigma}=\sum_{i=\rho,s} \frac{
\tilde{K}_{i}^{1/2}B_{i,\sigma}\tilde{\phi}_{i}^{+}
+ \tau \tilde{K}_{i}^{-1/2}D_{i,\sigma}\tilde{\phi}_{i}^{-} }{2}, \quad
\end{eqnarray}
where $\theta_{\sigma}$ is zero-mode phase satisfying $[ N_{\sigma},\theta_{\sigma} ] =i$.  Non-zero mode phase $\tilde{\phi}_{i}^{\pm}$ is given 
%%%%%%%%%%%%%%%%%%%%%%%%%%%%%%%%%%%%%%%%%%%%%%%%%%%%%%%%%%%%%%%%%%%%%%%%%%%%%%
\begin{eqnarray}
&&
\tilde{\phi}_{i}^{+}=\sum_{n=1}^{\infty}
\sqrt{\frac{2}{n}}\sin{k_{n}(x+d)} \left( 
\tilde{\alpha}_{i,n}+\tilde{\alpha}_{i,n}^{ \dagger}
 \right),  \\
 &&
\tilde{\phi}_{i}^{-}=\sum_{n=1}^{\infty}
\sqrt{\frac{2}{n}}
\frac{\cos{k_{n}(x+d)}}{i}
\left( 
\tilde{\alpha}_{i,n}- \tilde{\alpha}_{i,n}^{ \dagger}
 \right).  \quad
\end{eqnarray}
%%%%%%%%%%%%%%%%%%%%%%%%%%%%%%%%%%%%%%%%%%%%%%%%%%%%%%%%%%%%%%%%%%%%%%%%%%%%%%
The mode expansion of $\psi_{\sigma}^{\rm R,L}$ are similarly obtained.
Using the boundary operators, the tunnel Hamiltonian $H_{\rm T}=H_{\rm T}^{\rm R}+H_{\rm T}^{\rm L}$ is given 
%%%%%%%%%%%%%%%%%%%%%%%%%%%%%%%%%%%%%%%%%%%%%%%%%%%%%%%%%%%%%%%%%%%%%%%%%%%%%%
\begin{eqnarray}
H_{\rm T}^{\rm R}&=&\sum_{\sigma}t_{\sigma}^{\rm R}
\psi_{-,\sigma}^{\dagger}(d)\psi_{-,\sigma}^{\rm R}(d) +{\rm h.c.} , \\
H_{\rm T}^{\rm L}&=&\sum_{\sigma}t_{\sigma}^{\rm L}
\psi_{+,\sigma}^{\dagger}(-d)\psi_{+,\sigma}^{\rm L}(-d)  +{\rm h.c.} .
\end{eqnarray}
%%%%%%%%%%%%%%%%%%%%%%%%%%%%%%%%%%%%%%%%%%%%%%%%%%%%%%%%%%%%%%%%%%%%%%%%%%%%%%
The transition rate $P_{i \to f}^{\rm R(L)}$ from a initial state $|i\rangle \equiv |N_{\sigma},N_{-\sigma}\rangle$ to a final state $|f\rangle \equiv |N_{\sigma}+ q ,N_{-\sigma}\rangle$ ($q=\pm 1$), via tunneling processes at the right (left) end of the island, can be calculated perturbatively. 
To the lowest order, it becomes
%%%%%%%%%%%%%%%%%%%%%%%%%%%%%%%%%%%%%%%%%%%%%%%%%%%%%%%%%%%%%%%%%%%%%%%%%%%%
\begin{eqnarray}
P_{i \to f}^{\rm R (L)}=\int_{-\infty}^{\infty}dt \ 
\langle 
\langle i| H_{\rm T}^{\rm R (L)}(t) |f \rangle 
\langle f| H_{\rm T}^{\rm R (L)}(0) |i\rangle
\rangle, 
\end{eqnarray}
%%%%%%%%%%%%%%%%%%%%%%%%%%%%%%%%%%%%%%%%%%%%%%%%%%%%%%%%%%%%%%%%%%%%%%%%%%
where the time evolution of operators are given in interaction representation $A(t)=U^{\dagger}(t)AU(t)$ with $U=\exp[-it(H_{\rm zero}+H_{\rm fluc}+H_{\rm wires})]$.
We obtain the expression for $P_{i \to f}^{j}$ $(j={\rm R,L})$ as,
%%%%%%%%%%%%%%%%%%%%%%%%%%%%%%%%%%%%%%%%%%%%%%%%%%%%%%%%%%%%%%%%%%%%%%%%%%
\begin{eqnarray}
&&P_{i \to f}^{ j}= \frac{1}{\beta} \exp[-\frac{\beta \varepsilon_{\sigma}^{j}}{2}]   \gamma^{j}(\varepsilon_{\sigma}^{j},\beta), \label{eq:transition rate} \\
&&  \gamma^{j}(\varepsilon_{\sigma}^{j},\beta) \sim
\frac{1}{\pi}\left( \frac{t_{\sigma}^{j}}{v_{{\rm F},\sigma}} \right)^{2}
\left( \frac{\pi}{\beta \Lambda} \right)^{\frac{\lambda_{\sigma}}{2}-2} 
\prod_{i=\rho,s} \left(\frac{3\tilde{v}_{i}/d}{\Lambda}\right)^{\frac{D_{i,\sigma}^{2}}{2\tilde{K}_{i}}} \nonumber \\
&& \qquad \quad \times
2^{\lambda_{\sigma}-1} B\left[ 
\frac{\lambda_{\sigma}+i\beta \varepsilon_{\sigma}^{j}/\pi}{2}, \frac{\lambda_{\sigma}-i\beta \varepsilon_{\sigma}^{j}/\pi}{2}
 \right],  \label{eq:tunneling rate}
\end{eqnarray}
%%%%%%%%%%%%%%%%%%%%%%%%%%%%%%%%%%%%%%%%%%%%%%%%%%%%%%%%%%%%%%%%%%%%%%%%%%
where $\varepsilon_{\sigma}^{j} \equiv \langle f| H_{\rm zero} |f \rangle -\langle i| H_{\rm zero} |i \rangle -q \mu_{j}$ is the energy difference between the two states.
The line shape of the conductance near the peak is roughly given by the beta function in (\ref{eq:tunneling rate}), and the peak height scales like $\propto (T/\Lambda)^{\lambda_{\sigma}/2-2}$ as lowering temperature. 
The probability $P_{i}$ to find the system in $| i \rangle$, and the current of spin $\sigma$ electrons $I_{\sigma}$ are obtained by solving a set of master equations,
%%%%%%%%%%%%%%%%%%%%%%%%%%%%%%%%%%%%%%%%%%%%%%%%%%%%%%%%%%%%%%%%%%%%%%%%%%
\begin{eqnarray}
\frac{d}{dt}P_{i}&=&\sum_{i'} P_{i' \to i} P_{i'} - P_{i \to i'}P_{i}=0, \label{eq:master equation} \\
I_{\sigma} &=&e \sum_{i,q}
q P_{i} (P^{\rm R}_{i \to f}-P^{\rm L}_{i \to f} ), \label{eq:current by sequential tunneling}
\end{eqnarray}
%%%%%%%%%%%%%%%%%%%%%%%%%%%%%%%%%%%%%%%%%%%%%%%%%%%%%%%%%%%%%%%%%%%%%%%%%%
where $P_{i \to i'}=P_{i \to i'}^{\rm R}+P_{i \to i'}^{\rm L}$.

\begin{figure}[b]
\includegraphics[width=77.35mm, height=70.265mm]{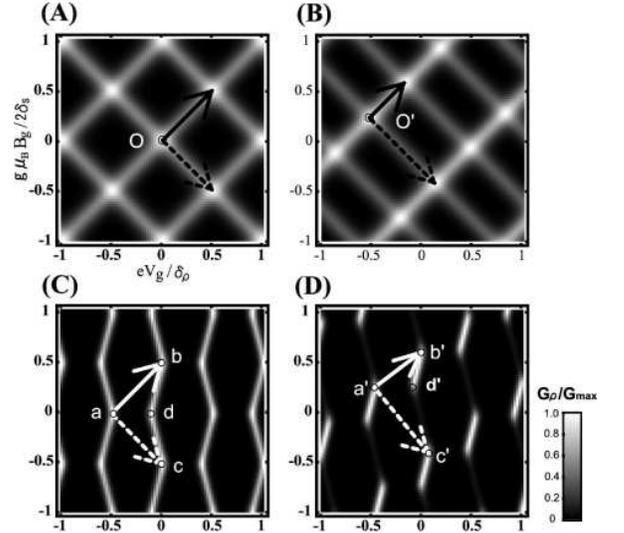} 
\vspace{-0.2cm}
\caption{\label{fig:weak link conductance} Zero bias conductance $G_{\rho}$ is plotted as a function of gate voltage and magnetic field normalized by $\delta_{i}=\frac{v_{i}}{K_{i}} \frac{\pi }{2d}$, with $T={10^{-1}} \frac{\pi v_{\rm F}}{4d}={10^{-3}} \Lambda$ and $(t_{\sigma}^{j}/v_{\rm F,\sigma})^2/\pi=10^{-2}$. Different values of interaction and magnetic field are taken for each figures; (A) $K_{\rho}=K_{s}=1$, $\Delta/v_{\rm F}=0$, (B) $K_{\rho}=K_{s}=1$, $\Delta/v_{\rm F}=0.3$, (C) $K_{\rho}=0.5$, $K_{s}=1$, $\Delta/v_{\rm F}=0$, and (D) $K_{\rho}=0.5$, $K_{s}=1$, $\Delta/v_{\rm F}=0.3$.}
\end{figure}

The zero bias conductance $G_{\rho}=G_{\uparrow}+G_{\downarrow}$ is evaluated from (\ref{eq:transition rate})-(\ref{eq:current by sequential tunneling}) and plotted as a function of gate voltage and gate magnetic field in Fig.\ref{fig:weak link conductance}.
%%%%%%%%%%%%%%%%%%%%%%%%%%%%%%%%%%%%%%%%%%%%%%%%%%%%%%%%%%%%%%%%%%%%%%%%%%
%%%%%%%%%%%%%%%%%%%%%%%%%%%%%%%%%%%%%%%%%%%%%%%%%%%%%%%%%%%%%%%%%%%%%%%%%%
The conductance peak form lattice in $(V_{\rm g},B_{\rm g})$ plane, and one can see the spin-charge mixing effect on the shape as well as in the case of weak barriers.
Arrows in figures are the primary unit vectors, and a translation by a solid (dotted) arrow corresponds to a change in the average particle number of down (up) spin electrons in the island by one.
When spin and charge excitations are degenerate ($K_{\rho}=K_{s}$), the primary unit vectors of the lattice pattern are in the direction of $\pm \frac{\pi}{4}$ independently of the magnetic field $\Delta$ as shown in Fig.\ref{fig:weak link conductance}(A) and Fig.\ref{fig:weak link conductance}(B).
On the other hand, when spin-charge separation does hold ($K_{\rho}, K_{s} \neq 1$), the vectors rotate as increasing $\Delta$ as seen in Fig.\ref{fig:weak link conductance}(C) and Fig.\ref{fig:weak link conductance}(D).
From Eq.(\ref{eq:zero mode 1}-\ref{eq:zero mode 2}), one can ensure that the angle of the solid (dotted) arrow $\vartheta_{\downarrow}$ ($\vartheta_{\uparrow}$) is given
%%%%%%%%%%%%%%%%%%%%%%%%%%%%%%%%%%%%%%%%%%%%%%%%%%%%%%%%%%%%%%%%%%%%%%
\begin{eqnarray}
\vartheta_{\sigma} =\tan^{-1}{\frac{-\sigma-K_{s}^{2} \Delta/v_{\rm F}}{1+\sigma K_{\rho}^{2} \Delta/v_{\rm F}}}.
\end{eqnarray}
%%%%%%%%%%%%%%%%%%%%%%%%%%%%%%%%%%%%%%%%%%%%%%%%%%%%%%%%%%%%%%%%%%%%%%
Thus the expansion of the rotation angle $\delta \vartheta$ to the linear order in $\Delta$, yields the same result as Eq.(\ref{eq: rotation angle}) obtained for the weak barrier case.
This could indicate that applying a magnetic field causes a rotation of the lattice of conductance peaks by the same angle for entire range of barrier strength to the linear order.
Of course, being similar to the weak barrier case, the splits of scaling dimensions for tunneling amplitudes between two spins results in strong suppression of the peak height along the resonant line $\overline{\rm d'c'}$ compared to $\overline{{\rm b'd'}}$ in Fig.\ref{fig:weak link conductance}(D). 
Another interesting result in the strong barrier limit is that for $K_{\rho}\neq K_{s}$ there are no resonance points of four number states (say $|n,m\rangle$, $|n+1,m\rangle$, $|n,m+1\rangle$ and $|n+1,m+1\rangle$) like O and ${\rm O}'$.
Instead, such a resonance point splits into two; one like a and ${\rm a}'$ where the three states $|n,m\rangle$, $|n+1,m\rangle$ and $|n,m+1\rangle$ degenerate, and the other like d and ${\rm d}'$ where $|n+1,m+1\rangle$, $|n+1,m\rangle$ and $|n,m+1\rangle$ do.
The length between the separated resonance points $\overline{\rm{ad}}$ and $\overline{\rm{a}'{\rm d}'}$, corresponds to $eV_{\rm g}=\frac{\pi v_{\rm F}}{4d}(K_{\rho}^{-2}-K_{s}^{-2})$.
 It can thus be a measure for the strength of spin-charge separation.
%%%%%%%%%%%%%%%%%%%%%%%%%%%%%%%%%%%%%%%%%%%%%%%%%%%%%%%%%%%%%%%%%%%%%%%
\section{Summary}
We have discussed the spin-charge mixing effect on the resonant tunneling in spin-polarized Tomonaga-Luttinger liquid under magnetic fields.
The zero bias conductance is calculated as a function of gate voltage $V_{{\rm g}}$ and gate magnetic field $B_{{\rm g}}$.
Conductance peaks form a lattice structure in $(V_{{\rm g}},B_{{\rm g}})$ plane.
We find two effects of the spin-charge mixing in the plot of zero bias conductance;
(i) The primary unit vectors of the lattice pattern rotate as increasing magnetic field due to ``spin-charge mixing''.
(ii) The amplitude of conductance oscillation differs significantly at low temperature between two spins, which originates from the split of the scaling dimension of impurity potential.
For systems with appropriate interaction parameters, the impurity potential can become a spin-filter that selects electrons of one spin orientation to pass through.
We should note that such spin-filtering phase appears only for the systems with spin-anisotropic interaction ($K_{s} \neq 1$), whose candidates in real systems haven't been found yet. 
However, recent studies predicts that spin-orbit interactions in a quantum wire, whose strength is tunable by the gate voltage, can also cause spin-charge mixing \cite{ref15} as the Zeeman effect does.
Moreover Gritsev $et \ al.$ \cite{ref16} show that an interaction parameter $K_{s}$ can be renormalized and shift from $K_{s}=1$ in the presence of Rashba coupling.
These facts may suggest a possibility to lead such spin-filtering phase to an experimentally accessible region.

Finally we would like to address the possibility for the experimental test of our theory.
If one uses armchair carbon nanotube as a TL wire, there arise two difficulties to observe the spin-charge mixing. 
Armchair nanotubew are described as four components TL liquid (spinful TL liquid with two bands $p=\pm$), having a symmetry in the band structure $v_{{\rm F},p}=v_{{\rm F},-p}$ at the Fermi level $\epsilon_{\rm F}=0$~\cite{egger&gogolin}.
Due to this unique band structure, the symmetry between up and down spins will not be broken even when applied magnetic field. Hense the spin-charge mixing does not occur for $\epsilon_{\rm F}=0$. This is one difficulty.
However, for doped carbon nanotubes ($\epsilon_{\rm F} \neq 0$) with the band structure symmetry broken $v_{{\rm F},p} \neq v_{{\rm F},-p}$, the spin-charge mixing occur when applied a strong magnetic field.
Another difficulty is on realization of such a strong magnetic fields that the spin-cahrge mixing effect can be measured.
In our case, a significant change in the Fermi velocity about $\frac{\Delta}{v_{\rm F}}=\frac{v_{{\rm F},\uparrow}-v_{{\rm F},\downarrow}}{2v_{\rm F}} \ge 0.1$ is needed.
Such situation may be difficult to prepare in carbon nanotubes with a band width  $t \sim 2.5 \ {\rm eV}$ \cite{ref17} since $g$-factor for electrons in carbon nanotubes is $g \sim 2$ \cite{ref18}, which means a magnetic field 1T amounts to $g \mu_{B} B \sim 0.12 \ {\rm meV}$.
To make velocity difference $\Delta/v_{\rm F} = 0.1$, it requires $B \sim 1.25 \times 10^{4}$ tesla in case of carbon nanotubes.
And the preparation of the local magnetic field $B_{\rm g}$ with a submicron meter scale is also an open issue, though there are some works reporting magnetic fields with a micron meter scale~\cite{ramadan}.
However, if one uses a quantum wire with a small Fermi energy and a large $g$-factor {\it e.g.} an InSb quantum wire with $g \sim -50$, two problems lying on carbon nanotubes are cleared, and our predictions can possibly be verified in experiments.
Ultracold fermionic atoms in optical lattices~\cite{ultracold}, which shows remarkable progress in experiments, with controllable parameters such as interaction parameters, lattice shapes, and the potential height, will also give us another conceivable stage to test our theory.

It is the more important issue of spintronics to suggest new ways of creating spin filters {\it i.e.} the way to modify the split of scaling dimensions between two spin channels in more realistic models.
However the concepts of ``spin-charge mixing effect'' will play one key role in this matter.

\acknowledgements
 We thank N. Yokoshi and T. Kimura for useful comments and discussions. This work is supported by The 21st Century COE Program (Holistic Research and Education center for Physics of Selforganization Systems) at Waseda University from the Ministry of Education, Sports, Culture, Science and Technology of Japan. K.K. is supported by the Japan Society for promotion of Science.

\appendix
\section{Luttinger parameters for spin-charge mixed system}
Luttinger parameters in Eq.(\ref{TL hamiltonian}) are given,
\begin{widetext}
%%%%%%%%%%%%%%%%%%%%%%%%%%%%%%%%%%%%%%%%%%%%%%%%%%%%%%%%%%%
\begin{eqnarray}
\tilde{v}_{\rho}^2&=& \frac{v_{\rho}^2+v_{s}^{2}+2 \Delta^2 }{2} +\frac{v_{\rho}^{2}-v_{s}^2}{2}
\sqrt{1+\frac{4 \Delta^2 (v_{\rho}K_{\rho}+v_{s}/K_{s})(v_{s}K_{s}+v_{\rho}/K_{\rho})}{(v_{\rho}^2-v_{s}^2)^2}}, \nonumber \\ 
\tilde{v}_{s}^2 &=& \frac{v_{\rho}^2+v_{s}^{2}+2 \Delta^2 }{2}- \frac{v_{\rho}^{2}-v_{s}^2}{2}
\sqrt{1+\frac{4 \Delta^2 (v_{\rho}K_{\rho}+v_{s}/K_{s})(v_{s}K_{s}+v_{\rho}/K_{\rho})}{(v_{\rho}^2-v_{s}^2)^2}}, \nonumber \\
%%%%%%%%%%%%%%%%%%%%%%%%%%%%%%%%%%%%%%%%%%%%%%%%%%%%%%
\tilde{K}_{\rho}&=&\sqrt{
\frac{v_{\rho}K_{\rho} \cos{\alpha}^2
+v_{s}K_{s}  (\frac{\sin{\alpha}}{y})^2
+\Delta \frac{\sin{2 \alpha}}{y}}
{\frac{v_{\rho}}{K_{\rho}} \cos{\alpha}^2
+\frac{v_{s}}{K_{s}} (y \sin{\alpha})^2
+\Delta y \sin{2 \alpha}}
} , \quad
%%%%%%%%%%%%%%%%%%%%%%%%%%%%%%%%%%%%%%%%%%%%%%%%%%%%%%%%%
\tilde{K}_{s}=\sqrt{
\frac{v_{s}K_{s} \cos{\alpha}^2
+v_{\rho}K_{\rho}  (y \sin{\alpha})^2
-\Delta y \sin{2 \alpha}}
{\frac{v_{s}}{K_{s}} \cos{\alpha}^2
+\frac{v_{\rho}}{K_{\rho}} (\frac{\sin{\alpha}}{y})^2
-\Delta \frac{\sin{2 \alpha}}{y}}
}, \nonumber 
\end{eqnarray}
%%%%%%%%%%%%%%%%%%%%%%%%%%%%%%%%%%%%%%%%%%%%%%%%%%%%%%%%%%%%%%
\end{widetext}
where $\alpha$ and $y$ are given in Eq.(\ref{rotation angle}).
%
%%%%%%%%%%%%%%%%%%%%%%%%%%%%%%%%%%%%%%%%%%%%%%%%%%%%%%%%%%%%
\section{Linear transformation}
%%%%%%%%%%%%%%%%%%%%%%%%%%%%%%%%%%%%%%%%%%%%%%%%%%%%%%%%%%%%
Here we represent the linear transformation in Eq.(\ref{linear transform}) in terms of ladder operators of TL bosons $\alpha_{i}(k)$, $\alpha_{i}^{\dagger}(-k)$, $\tilde{\alpha}_{i}(k)$ and $\tilde{\alpha}_{i}^{\dagger}(-k)$, which diagonalize TL Hamiltonian as $H_{\rm TL}=\sum_{k,i}\tilde{v}_{i}|k| \left( \tilde{\alpha}_{i}^{\dagger}(k)\tilde{\alpha}_{i}(k)+1/2 \right)$.
The phase variables $\phi_{i}$ and $\Pi_{i}$ are expanded in terms of the ladder operators as
%%%%%%%%%%%%%%%%%%%%%%%%%%%%%%%%%%%%%%%%%%%%%%%%%%%%%%%%%%%%%%%%
\begin{eqnarray}
\phi_{i} &=&\sum_{k} \sqrt{\frac{\pi K_{i}}{2|k|L}} \left( \alpha_{i}(k)+\alpha_{i}^{\dagger}(-k) \right) e^{ikx}, \label{mode expansion 1} \\ 
\Pi_{i} &=& \frac{1}{i} \sum_{k} \sqrt{\frac{|k| }{2 \pi K_{i} L}} \left( \alpha_{i}(k)-\alpha_{i}^{\dagger}(-k) \right) e^{ikx}. \label{mode expansion 2}
\end{eqnarray}
%%%%%%%%%%%%%%%%%%%%%%%%%%%%%%%%%%%%%%%%%%%%%%%%%%%%%%%%%%%%%
The representation for $\tilde{\phi}_{i}$ and $\tilde{\Pi}_{i}$ are obtained similarly by putting $K_{i},\alpha_{i}\to \tilde{K}_{i},\tilde{\alpha}_{i}$ into the above equations.
Focusing on a $k$-component here, we omit the index $k$ for simplicity.
From the above expressions, the transformation in Eq.(\ref{linear transform}) is expressed as ${\mathcal A} = {\mathcal W} \tilde{ \mathcal A}$ with ${\mathcal A}=(\alpha_{\rho},\alpha_{s},\alpha_{\rho}^{\dagger},\alpha_{s}^{\dagger})^{T}$, $ \tilde{\mathcal A}=(\tilde{\alpha}_{\rho},\tilde{\alpha}_{s},\tilde{\alpha}_{\rho}^{\dagger}, \tilde{\alpha}_{s}^{\dagger})^{T}$, and  
a 4$\times$4 matrix ${\mathcal W}$;
%%%%%%%%%%%%%%%%%%%%%%%%%%%%%%%%%%%%%%%%%%%%%%%%%%%%%%%%%%%%%%%%
\begin{eqnarray}
&& {\mathcal W}= \frac{1}{2} \left(
\begin{array}{cc}
P_{\phi}+P_{\Pi} & P_{\phi}-P_{\Pi} \\
P_{\phi}-P_{\Pi} & P_{\phi}+P_{\Pi} 
\end{array}
\right),  \\
&& P_{\phi}= \left(
\begin{array}{cc}
\cos \alpha  \sqrt{\tilde{K}_{\rho}/K_{\rho}} & 
- \frac{1}{y}\sin \alpha \sqrt{\tilde{K}_{s}/K_{\rho}} \\
y \sin \alpha \sqrt{\tilde{K}_{\rho}/K_{s}}  & 
\cos \alpha \sqrt{\tilde{K}_{s}/K_{s}} 
\end{array}
\right), \nonumber \\ 
&& P_{\Pi} = \left(
\begin{array}{cc}
\cos \alpha \sqrt{K_{\rho}/\tilde{K}_{\rho}} & 
-y \sin \alpha \sqrt{K_{\rho}/\tilde{K}_{s}}\\
\frac{1}{y}\sin \alpha \sqrt{K_{s}/\tilde{K}_{\rho}} &
 \cos \alpha \sqrt{K_{s}/\tilde{K}_{s}}
\end{array}
\right). \nonumber 
\end{eqnarray}
%%%%%%%%%%%%%%%%%%%%%%%%%%%%%%%%%%%%%%%%%%%%%%%%%%%%%%%%%%%%%%%%
For sure, it can be easily checked that the transformation matrix ${\mathcal W}$ satisfies a normalization condition for a Bogoliubov transformation ${\mathcal W}{\mathcal C}{\mathcal W}^{\dagger}={\mathcal C}$ with ${\mathcal C}={\rm diag}(1,1,-1,-1)$.
%
%
%
%
%%%%%%%%%%%%%%%%%%%%%%%%%%%%%%%%%%%%%%%%%%%%%%%%%%%%%%%%%%%%%%%%%%%%%%%
\section{The effect of noninteracting reservoir}
%%%%%%%%%%%%%%%%%%%%%%%%%%%%%%%%%%%%%%%%%%%%%%%%%%%%%%%%%%%%%%%%%%%%%%%
%%%%%%%%%%%%%%%%%%%%%%%%%%%%%%%%%%%%%%%%%%%%%%%%%%%%%%%%%%%%%%%%%%%%%%%
Here we calculate the zero bias conductance of a clean TL wire under magnetic field, connected to Fermi liquid reservoirs. As previous works~\cite{ref12,ref13,ref14} have shown, the scaling factor of conductance should disappear by the effect of reservoirs. We check whether their statements can also be applied to a polarized TL liquid, and whether our result in Eq.(\ref{eq: conductance in clean limit 2}), which implies that applying bias voltage generates spin current by spin-charge mixing effect, is an artifact of the assumption to be an infinite system.
We consider a polarized TL wire is connected to reservoirs of noninteracting Fermi liquid at $x=\pm d$.

From Hamiltonian (\ref{hamiltonian before diagonalized}), the imaginary time action becomes,
%%%%%%%%%%%%%%%%%%%%%%%%%%%%%%%%%%%%%%%%%%%%%%%%%%%%%%%%%%%%%%%%%%%%%%%%%%%%
\begin{eqnarray}
S   &=&\frac{ 1} {2 \pi}\int_{0}^{\beta}d\tau \int_{-\infty}^{\infty} dx \ 
\vec{\phi}^{T}  \hat{M}  \vec{\phi} \ ,  \\
%%%%%%%%%%%%%%%%%%%%%%%%%%%%%%%%%%%%%%%%%%%%%%%%
\vec{\phi}&=&
\left(
\begin{array}{cc}
\phi_{\rho} & \phi_{s}
\end{array}
\right)^{T},
\nonumber \\
%%%%%%%%%%%%%%%%%%%%%%%%%%%%%%%%%%%%%%%%%%%%%%%%
\hat{M} &=&
\left(
\begin{array}{cc}
v_{\rho}K_{\rho}& \Delta \\
\Delta & v_{s}K_{s} 
\end{array}
\right)^{-1} \partial_{\tau}^{2} +
\partial_{x}
\left(
\begin{array}{cc}
\frac{v_{\rho}}{K_{\rho}} & \Delta \\
\Delta & \frac{v_{s}}{K_{s}} 
\end{array}
\right)
\partial_{x}.  \nonumber
\end{eqnarray}
%%%%%%%%%%%%%%%%%%%%%%%%%%%%%%%%%%%%%%%%%%%%%%%%%%%%%%%%%%%%%%%%%%%%%%%%%%
$\Delta$, $v_{i}$ and $K_{i}$ are $x$-dependent parameters and abruptly change at the boundaries,
%%%%%%%%%%%%%%%%%%%%%%%%%%%%%%%%%%%%%%%%%%%%%%%%%%%%%%%%%
\begin{eqnarray}
&&v_{i}(x)=\left\{
\begin{array}{cc}
v_{i} & -d<x<d \\
v_{\rm F} & {\rm otherwise,}
\end{array}
\right.
%%%%%%%%%%%%%%%%%%%%%%%%%
 \ K_{i}(x)=\left\{
\begin{array}{cc}
K_{i} & -d<x<d \\
1 & {\rm otherwise,}
\end{array}
\right.
\nonumber \\
%%%%%%%%%%%%%%%%%%%%%%%%%
&&\Delta(x)=\left\{
\begin{array}{cc}
\Delta & -d<x<d \\
0 & {\rm otherwise.}
\end{array}
\right. 
\end{eqnarray}
%%%%%%%%%%%%%%%%%%%%%%%%%%%%%%%%%%%%%%%%%%%%%%%%%%%%%%
Conductance can be calculated from the Green's function of bosonic field, 
%%%%%%%%%%%%%%%%%%%%%%%%%%%%%%%%%%%%%%%%%%%%%%%%%%%%%%%%%%%%%%%%%%%%%%%%%%%%%
\begin{eqnarray}
&& \hat{G}(\tau, x,y)= \sum_{\omega_{n}}   
\hat{G}_{\omega_{n}}(x,y)  e^{i \omega_{n} \tau }  
\label{Green's function} \\ 
&& =\left(
\begin{array}{cc}
\left< T_{\tau} \phi_{\rho}(\tau,x)\phi_{\rho}(0,y) \right> &
\left< T_{\tau} \phi_{\rho}(\tau,x)\phi_{s}(0,y) \right> \\
\left< T_{\tau} \phi_{s}(\tau,x)\phi_{\rho}(0,y) \right> &
\left< T_{\tau} \phi_{s}(\tau,x)\phi_{s}(0,y) \right>
\end{array}
\right) ,
\nonumber
\end{eqnarray}
%%%%%%%%%%%%%%%%%%%%%%%%%%%%%%%%%%%%%%%%%%%%%%%%%%%%%%%%%%%%%%%%%%%%%%%%%%%%%
which satisfies the equation,
%%%%%%%%%%%%%%%%%%%%%%%%%%%%%%%%%%%%%%%%%%%%%%%%%%%%%%%%%%%%%%%%%%%%%%%%%%%%%
\begin{eqnarray}
\hat{M}_{\omega_{n}}  \hat{G}_{\omega_{n}}(x,y)=\hat{1} \cdot \delta (x-y) . 
\label{eq: equation for greens function}
\end{eqnarray}
%%%%%%%%%%%%%%%%%%%%%%%%%%%%%%%%%%%%%%%%%%%%%%%%%%%%%%%%%%%%%%%%%%%%%%%%%%%%%
Here $\hat{M}_{\omega_{n}}$ is Fourier component of $\hat{M}$ obtained by $\partial_{\tau} \to i\omega_{n}$.
DC charge current ($I_{\rho}$) and spin current ($I_{s}$) induced by a time independent electronic field $E(y)$ in TL wire $-d<y<d$, are determined from Kubo formula for $i=\rho,s$
%%%%%%%%%%%%%%%%%%%%%%%%%%%%%%%%%%%%%%%%%%%%%%%%%%%%%%%%%%%%%%%%%%%%%%%%%%%%%
\begin{eqnarray}
I_{i}(x)=\int_{-d}^{d} dy 
\lim_{\omega_{n} \to 0}
\left(
 - \omega_{n} \frac{e^2}{\pi} 
\left[
\hat{G}_{\omega_{n}} (x,y) \right]_{i,\rho}
\right)
E(y) .  \quad
\end{eqnarray}
%%%%%%%%%%%%%%%%%%%%%%%%%%%%%%%%%%%%%%%%%%%%%%%%%%%%%%%%%%%%%%%%%%%%%%%%%%%%%
We must solve (\ref{eq: equation for greens function}) under the boundary conditions at $x=\pm d,y$; (i) The Green's function should be continuous and (ii) the integration should satisfy
%%%%%%%%%%%%%%%%%%%%%%%%%%%%%%%%%%%%%%%%%%%%%%%%%%%%%%%%%%%%%%%%%%%%%%%%%%%%%
\begin{eqnarray}
&& \int_{-d-0}^{-d+0}dx \hat{M}_{\omega_{n}}  \hat{G}_{\omega_{n}}(x,y)=\int_{d-0}^{d+0}dx \hat{M}_{\omega_{n}}  \hat{G}_{\omega_{n}}(x,y)=\hat{0}, \nonumber \\
&& \int_{y-0}^{y+0}dx \hat{M}_{\omega_{n}}  \hat{G}_{\omega_{n}}(x,y)=\hat{1}.
\end{eqnarray}
%%%%%%%%%%%%%%%%%%%%%%%%%%%%%%%%%%%%%%%%%%%%%%%%%%%%%%%%%%%%%%%%%%%%%%%%%%%%%
One can find the solution to become
%%%%%%%%%%%%%%%%%%%%%%%%%%%%%%%%%%%%%%%%%%%%%%%%%%%%%%%%%%%%%%%%%
\begin{eqnarray}
\hat{G}_{\omega_{n}}(x,y)=
\Bigl(
\begin{array}{cc}
-\frac{1}{\omega_{n}} & 0 \\
0 & -\frac{1}{\omega_{n}}
\end{array}
\Bigr) 
+{\mathcal O}( \omega_{n}^{0} ).
\end{eqnarray}
%%%%%%%%%%%%%%%%%%%%%%%%%%%%%%%%%%%%%%%%%%%%%%%%%%%%%%%%%%%%%%%%%%%%%%%%%%%
Thus the conductance is given by $G_{\rho}^{(0)}=\frac{e^2}{\pi}$ and $G_{s}^{(0)}=0$ instead of Eq.(\ref{eq: conductance in clean limit 1}-\ref{eq: conductance in clean limit 2}). 
The result shows the conductance quantization to $\frac{e^2}{2 \pi}$ par spin channel also holds for spin-charge mixed systems under the magnetic field, as far as concerned the DC limit $ |\omega_{n}| \ll v_{\rm F}/2d $.
Moreover we can ensure this conclusion is unchanged when a magnetic field is also applied to reservoirs besides the one-dimensional region.

\newpage 
%Just because of unusual number of tables stacked at end


\end{document}